\documentclass[reprint,preprintnumbers, amsmath,amssymb,aps,prl,]{revtex4-1}
\usepackage{graphicx}% Include figure files
\usepackage{dcolumn}% Align table columns on decimal point
\usepackage{bm}% bold math
\usepackage{hyperref}% add hypertext capabilities
\begin{document}

%\preprint{APS/123-QED}

\title{Unexpected Effectiveness of Berezinskii-Kosterlitz-Thouless Description of Electron Paramagnetic Resonance Linewidth Behaviour in the 3-Dimensional Manganite Bi\textsubscript{0.5}Sr\textsubscript{0.5}Mn\textsubscript{0.9}Cr\textsubscript{0.1}O\textsubscript{3}}% Force line breaks with \\

\author{A. Ashoka\textsuperscript{$\dagger$}}%A 
\altaffiliation[Current Address:]{Cavendish Laboratory, University of Cambridge, J. J. Thomson Avenue, Cambridge CB3 0HE, United Kingdom}
\author{K. S. Bhagyashree}
\author{S.V. Bhat}

 
\affiliation{%
 Department of Physics, Indian Institute of Science, Bangalore-560012, India\\
 $\dagger$Academies' Summer Fellow
}%


\begin{abstract}

We find that the temperature dependence of the Electron Paramagnetic Resonance (EPR) linewidth observed in Bi\textsubscript{0.5}Sr\textsubscript{0.5}Mn\textsubscript{0.9}Cr\textsubscript{0.1}O\textsubscript{3} (BSMCO), as well as in certain other 3-dimensional manganites undergoing antiferromagnetic transitions is satisfactorily described by the Berezinskii-Kosterlitz-Thouless (BKT) model. We understand this unexpected result in terms of an effective 2-dimensional XY easy plane anisotropy induced by the magnetic field applied in the EPR experiment. This conclusion is supported by the field dependence of the BKT correlations observed in the quasi two-dimensional antiferromagnetic compound BaNi\textsubscript{2}V\textsubscript{2}O\textsubscript{8}.

\end{abstract}

\pacs{Valid PACS appear here}% PACS, the Physics and Astronomy
\maketitle

%\tableofcontents

Ideal two-dimensional (2D) Heisenberg magnets lack long range magnetic order\cite{Mermin}. However, the XY model with spins confined to a plane shows a topological phase transition at a finite temperature corresponding to binding and unbinding of vortices \cite{Berezinskii,Kosterlitz1973}.  While experimental evidence for such Berezinskii-Kosterlitz-Thouless (BKT) transitions was found in \textsuperscript{4}He superfluid films and superconducting films \cite{Goldman2013}, in condensed matter systems it has been difficult to observe a BKT transition. In the latter, even weak interlayer coupling that is invariably present leads to a long-range order, pre-empting the BKT transition in most cases. Above the long-range ordering temperature however, BKT signatures are still discernible as a characteristic exponential temperature dependence of the coherence length of the fluctuations. This was observed, for example, in quasi 2D materials such as BaNi\textsubscript{2}V\textsubscript{2}O\textsubscript{8} and more recently in certain chromium spinals which are nominally 3D systems but where geometric antiferromagnetic frustration is understood to have resulted in the reduction of effective dimensionality \cite{Heinrich2003,Hemmida2017}.  In this letter we report Electron Paramagnetic Resonance (EPR) studies which indicate that the Cr\textsuperscript{3+} doped bismuth strontium manganite Bi\textsubscript{0.5}Sr\textsubscript{0.5}Mn\textsubscript{0.9}Cr\textsubscript{0.1}O\textsubscript{3} (BSMCO) exhibits BKT like correlations even in the absence of 2-dimensionality and frustration of structural origin.  We explain this result in terms of frustration originating in the coexistence of antiferromagnetism and ferromagnetism intrinsic to doped manganites and 2-dimensionality induced by an applied magnetic field. We point out that EPR linewidth studies reported earlier on BaNi\textsubscript{2}V\textsubscript{2}O\textsubscript{8} in \cite{Heinrich2003,Waibel} provide additional support to magnetic field dependence of the BKT transition temperature. 

BSMCO belongs to the family of mixed valent manganites of the type ReAMnO\textsubscript{3} where Re is a trivalent rare earth ion such as La\textsuperscript{3+}, Nd\textsuperscript{3}+, Pr\textsuperscript{3+} etc. or Bi\textsuperscript{3+}, and A is a divalent alkaline earth ion such as Ca\textsuperscript{2+}, Sr\textsuperscript{2+} etc. In manganites there is a strong coupling between charge, spin and orbital degrees of freedom leading to exotic properties like colossal magnetoresistance, charge order and phase coexistence. They exhibit complex phase diagrams with fragile phase boundaries with structural, transport and magnetic properties extremely sensitive to the amount and nature of doping. Most manganites are intrinsically inhomogeneous, due to the presence of strong tendencies towards phase separation consisting of ferromagnetic (FM) metallic and antiferromagnetic (AF) insulating domains \cite{Dagotto}.  An enormous amount of experimental and theoretical work on manganites has accumulated over the last couple of decades though a consistent understanding of the experimental results  is yet to emerge. 

Spin being an important degree of freedom, EPR, in addition to magnetization studies, has been one of the useful experimental techniques for the  study of manganites. The EPR linewidth $\Delta H$,  which gives information on spin dynamics is the most important EPR parameter for study \cite{Huber2007}. A variety of $\Delta H(T)$ behaviour has been observed in different manganite systems and several different mechanisms such as spin only, spin-phonon and bottle necked relaxation have been invoked to explain the results; here we focus on systems showing AF transition, of which BSMCO is an example, where, as T is decreased towards $T_N$, $\Delta H$ continuously increases, diverging around $T_N$.
Such a behaviour of $\Delta H(T)$ was observed in the early report of Granado et al., in Ca\textsubscript{1-x}La\textsubscript{x}MnO\textsubscript{3} (x=0, 0.02 and 0.05) \cite{Granado2001}. They report that even the very low level of doping has `dramatic' effects on the EPR linewidth behaviour and therefore on the exchange mechanisms. They report difficulty in fitting their data to the commonly used Ginzberg-Landau critical model \cite{Oleaga}, according to which,
\begin{equation}
    \Delta H(T) = \frac{C}{(\frac{T}{T_c} - 1)^p}  + mT + H_0
\end{equation}
where $T_c$ is the Neel or Curie Temperature, $C$ is a proportionality constant and $p$ is the critical exponent that depends on the underlying spin and spatial degrees of freedom and theoretically takes values between 0.6 and 5.6 for a 3D Ising ferromagnet and 2D Ising ferromagnet respectively \cite{Benner}. A term linear in $T$ and a temperature independent term are added in Eq. (1) to describe the physics far away from the transition \cite{Heinrich2003}. Granado et al., find that eqn. (1) fits the data for the x = 0.05 sample satisfactorily, but it cannot fit the data for the x = 0 sample.  

To solve this difficulty with the critical model for CaMnO\textsubscript{3}, Granado et al. have suggested the use of a model put forward by Bhagat et al., that describes an exponential dependence of $\Delta H$ on T based on spin-freezing that was found to fit the CaMnO\textsubscript{3} data with $T_N$ at the magnetisation Neel temperature  \cite{Granado2001,Bhagat1981}. Bhagat et al. postulate that the spin relaxation rate is proportional to the strength of the frozen moment seen by the resonating spins, resulting in $\Delta H(T) \sim \int_0^{t_0} e^{\frac{-t}{\tau}} dt $, where $\tau$ is the relaxation time and $t_0$ is some characteristic time of the ESR experiment. Taking $\tau \propto (T-T_S)^{-1}$ leads to an exponential dependence of the linewidth function with respect to temperature \cite{Bhagat1981}, 
\begin{equation}
    \Delta H(T) = A \exp\bigg[-\frac{(T-T_S)}{T_0}\bigg] + mT + H_0
\end{equation}
where $A$ is a constant of proportionality, $T_S$ is the critical transition temperature for e.g. $T_N$ and $T_0$ is an empirical constant.

In recent work Hemmida et al. \cite{Hemmida2017}, report that $\Delta H(T)$ in 3D chromium spinels can be explained by the manifestly 2-dimensional BKT scenario which predicts the presence of a topological phase transition based on the binding and unbinding of vortices in  the 2D XY model. This gives the temperature dependence of the correlation length as \cite{Kosterlitz1973},
\begin{equation}
    \xi = \xi_0 \exp\bigg[ \frac{b}{(\frac{T}{T_{BKT}} - 1)^{0.5}}\bigg] 
\end{equation}
where, $T_{BKT}$ is the BKT transition temperature, $\xi_0$ is the infinite temperature correlation length, and b takes the value of $\pi /2$ for a square lattice \cite{Kosterlitz1973}, but has been theoretically shown to take an arbitrary value \cite{Kawamura2010}. We use the square lattice value of $b$ for all calculations and fits as we are dealing with perovskites. In general an ESR experiment probes the dynamic structure factor at approximately zero momentum $\textbf{q}$ (microwave radiation) \cite{Hemmida2017}. Following Benner and Boucher \cite{Benner}, assuming that the average vortex velocity $\bar{u}$ is temperature independent, and with $\gamma=\sqrt{\pi} \bar{u} / 2\xi $ we have $\Delta H \propto S_{\textbf{xx}}(\textbf{q}\rightarrow0,\omega \rightarrow 0) \propto \xi^2/\gamma \propto \xi^3$. Using Eq. (3), we then get,
\begin{equation}
    \Delta H(T) = \Delta H_{\infty} \exp\bigg[ \frac{3b}{\sqrt{(\frac{T}{T_{BKT}} - 1)}}\bigg] + mT + H_0
\end{equation}
Layered magnets with strong in-plane coupling J and a weak inter plane coupling J' giving rise to quasi two-dimensionality for the spin degrees of freedom have been studied to look for realisation of the BKT transition \cite{Bramwell1993,Heinrich2003}. It has been shown that only fluctuations on length scales less than the order of $L_{eff} =\sqrt{(J/J')}$ are two-dimensional in these layered magnets \cite{Bramwell1993} which provides a measure of planar anisotropy as developed by Bramwell and Holdsworth. They suggest the following expression for systems with weak in-plane anisotropy and inter plane coupling,
\begin{equation}
    \frac{J}{J'} = \exp \bigg[\frac{2b}{\sqrt{\frac{T_N}{T_{BKT}}-1}}\bigg]
\end{equation}
In typical weakly anisotropic layered 2D Heisenberg magnets, $J/J'$ is in the range of $10^3$ - $10^4$, as has been experimentally determined for known layered magnets \cite{Heinrich2003}.
\begin{figure}
    \centering
    \includegraphics[width=0.45\textwidth]{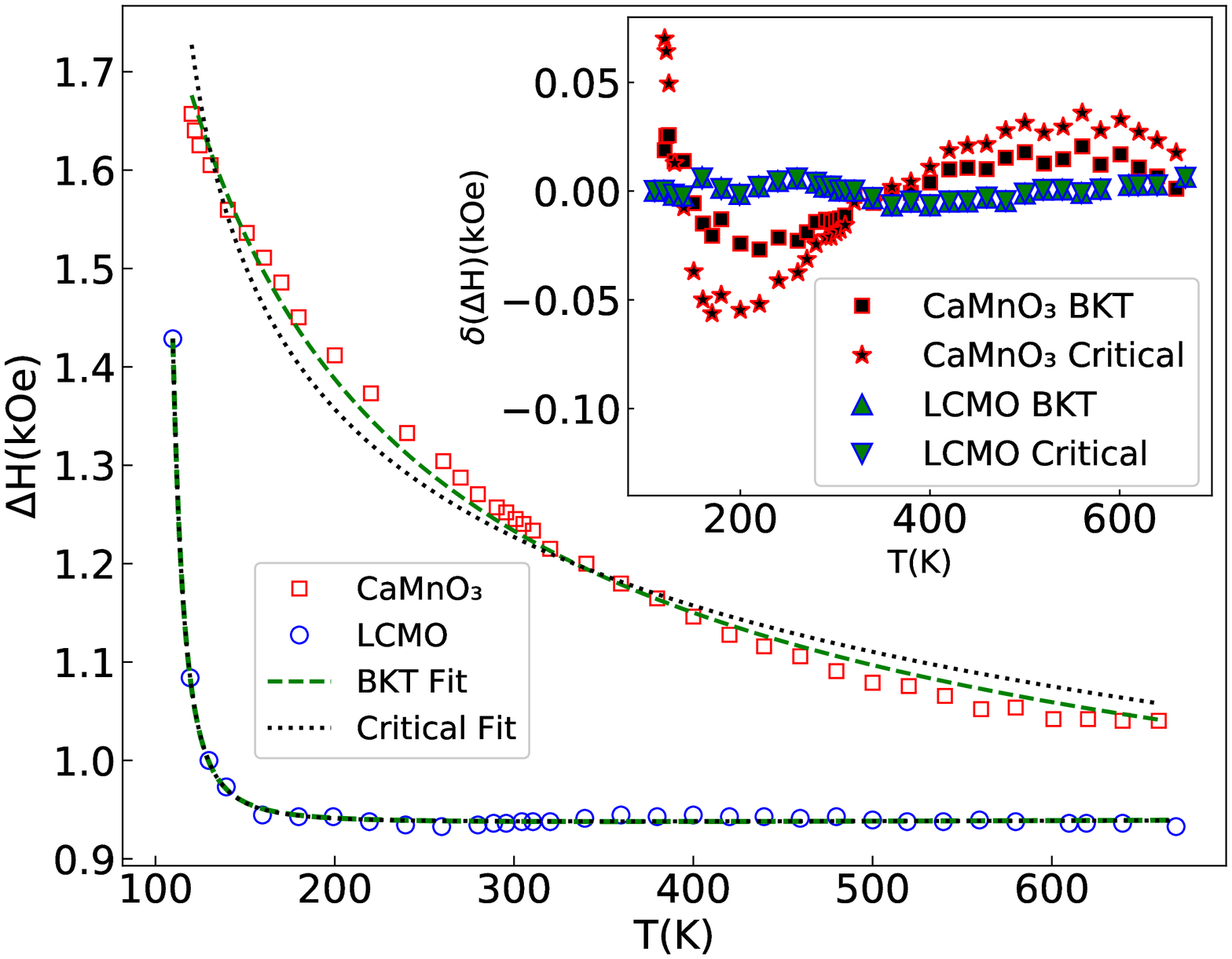}
    \caption{BKT and Critical Model Fits for LCMO and CaMnO\textsubscript{3}. Inset: Residuals from BKT and Critical fits of LCMO and CaMnO\textsubscript{3} data.}
\end{figure}

Quantum monte carlo simulations of AFM spins on a 2D square lattice with weak planar anisotropy have displayed evidence of a crossover from high temperature isotropic behaviour to two dimensional XY behaviour \cite{Cuccoli}. There is some evidence for a BKT transition and BKT like features to be induced at a finite temperature in an isotropic Heisenberg 2D AFM by the application of an external magnetic field, which allows for genuine XY behaviour in an extended temperature range \cite{Cuccoli2003,Cuccoli}. The magnitude of the applied field has a significant impact on the crossover temperature in these quasi two dimensional compounds and the relation as predicted by renormalisation group techniques \cite{Khokhlacev,Irkhin1999} is given by \cite{Cuccoli2003} as,

\begin{equation}
    t_{BKT} \simeq \frac{4\pi \rho_s /J}{\ln (A/h^2)}
\end{equation}

where A is a constant, J is the spin-spin coupling constant, $t = T/J$ is the reduced temperature, $\rho_S$ is the spin stiffness and $h=g\mu_BH/(JS)$ is the reduced magnetic field where $S=\sqrt{S(S+1)}$. 

\begin{figure}
    \centering
    \includegraphics[width=0.45\textwidth]{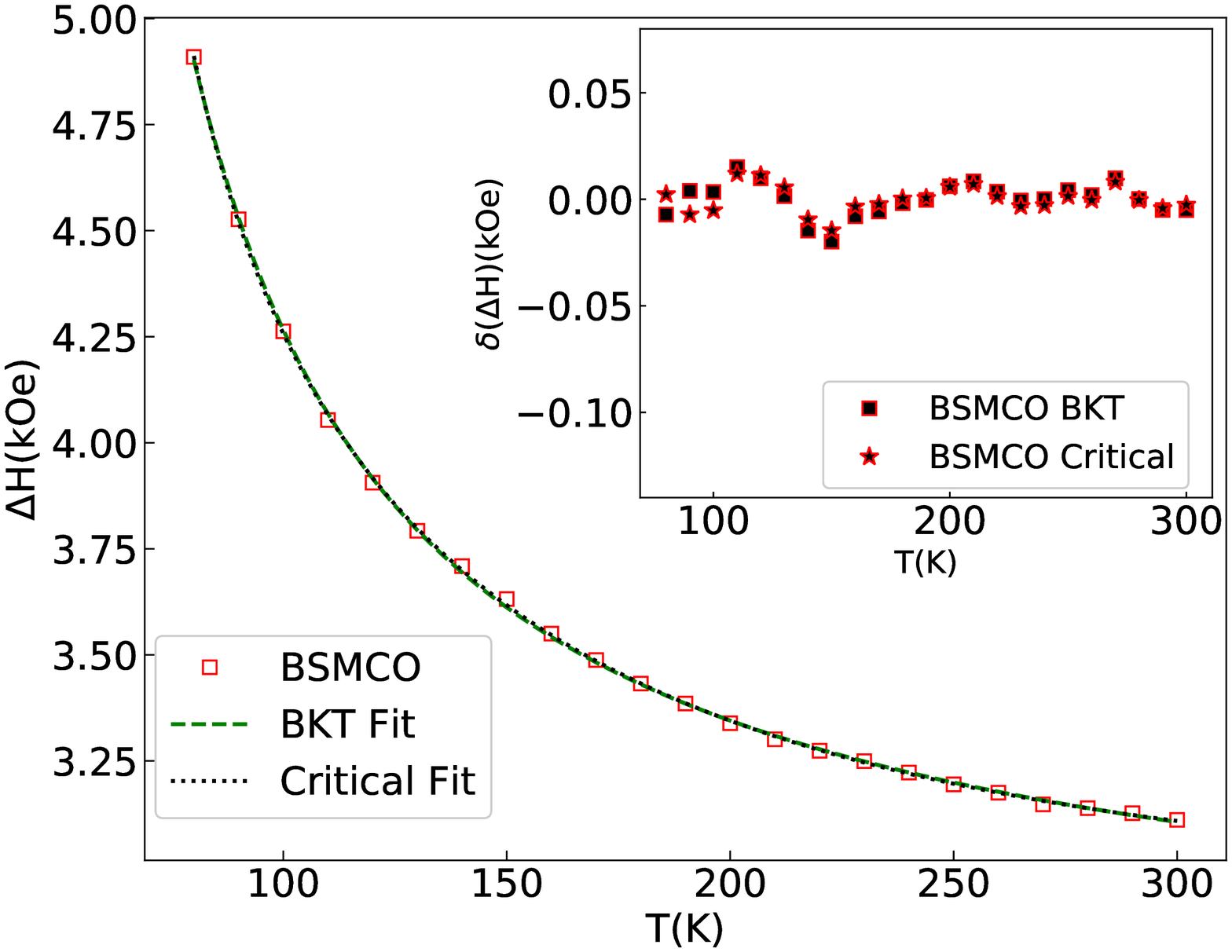}
    \caption{BKT and Critical Model Fits for BSMCO. Inset: Residuals from BKT and Critical fits of BSMCO Data.}
\end{figure}

We present our analysis of the results of Granado et al., on CaMnO\textsubscript{3} and La\textsubscript{0.05}Ca\textsubscript{0.95}MnO\textsubscript{3} in Figure 1\cite{Granado2001}. Graphical data of $\Delta H(T)$ for CaMnO\textsubscript{3} and La\textsubscript{0.05}Ca\textsubscript{0.95}MnO\textsubscript{3} (LCMO) was digitised and the errors in the digitised data were calculated to be $\delta (\Delta H)\simeq2.9 Oe$ and $\delta T \simeq 2 K$, both within the errors in the original data implying an ideal or near ideal reproduction of the published data through digitisation. Our results and analysis on BSMCO are presented in Figure 2. Preparation of polycrystalline powders of bulk BSMCO and subsequent structural and magnetic characterisation are reported by Bhagyashree et al. in \cite{Bhagyashree}. Linewidth of the signal has been determined by numerically fitting the data as described in \cite{Bhagyashree} to the broad lorentzian lineshape function which describes the data well until $\sim$ 70 K. The fit parameters for the critical, BKT and exponential models for all three samples investigated are summarised in Table 1 along with the coefficient of determination ($R^2$ value) which measures the goodness of fit.

\begin{table}[h!]
    \centering
\begin{tabular}{|p{1.8cm}||p{2cm}|p{2cm}|p{2cm}|}
 \hline
  Parameter From Fit  & CaMn0\textsubscript{3} &  LCMO & BSMCO \\
 \hline
 \multicolumn{4}{|c|}{Critical Model} \\
 \hline
 $C$ (kOe) & 1.35 & 0.0052 & 3.846 \\
 $p$ & 0.14 & 2.82 & 0.24 \\
 $T_N$* (K) & 102.0 & 92.4 & 56. 9\\
 $m$ (kOe/K) & 0.0 & 2.1 $\times$10\textsuperscript{-6}& 1.4 $\times$10\textsuperscript{-3} \\
 $H_0$ (kOe) & 0.00 & 0.93 & 2.2$\times$10\textsuperscript{-5}  \\
 $R^2$ & 0.98489 & 0.99925 & 0.99912 \\
 \hline
 \multicolumn{4}{|c|}{BKT Model} \\
 \hline
 $H_{\infty}$ (kOe) & 0.73 & 0.0001 & 0.28\\
 $b$* & 1.570 & 1.572 & 1.572 \\
 $T_{BKT}$ (K) & 3.5 & 83.0 & 14.7\\
 $m$ (kOe/K) & 0.0 & 7.5$\times$10\textsuperscript{-6} & 0.0 \\
 $H_0$ (kOe) & 0.00 & 0.93 & 2.29 \\
 $R^2$ & 0.99689 & 0.99918 & 0.99986\\
 \hline
 \multicolumn{4}{|c|}{Exponential (Spin Freezing Model)} \\
 \hline
 $A$ (kOe) & 0.67 & 6.54 & 0.24 \\
 $T_0$ (K) & 186.3 & 8.60 & 55.5\\
 $T_N$* (K) & 111.3 & 87.3 & 102.1\\
 $m$ (kOe/K) & 0 & 0 & 0\\
 $H_0$ (kOe) & 2$\times$10\textsuperscript{-6} & 0.94 & 3.12 \\
 $R^2$ & 0.99959 & 0.99782 & 0.99752 \\
\hline
\end{tabular}
 \caption{Summary of fit parameters and goodness of fit for each model. The starred parameters were constrained to within reasonable range of physically realistic or experimentally known values.}
\end{table}

It is clear from Figure 1 that applying the critical model, i.e, Eq. (1) is not satisfactory in fitting the experimental data for CaMnO\textsubscript{3} as reported by Granando et al. and the coefficient of determination of the BKT and exponential model indicate that both outperform the critical model for CaMnO\textsubscript{3}. The extended temperature range ($\sim$ 600K) over which the transition takes place and the large residuals for the critical model fit at $T<200K$ in Figure 1 suggest that the critical model, which considers fluctuations only close to the Neel temperature, is not appropriate. 

We suggest that the exponential model's good fit with $T_N$ at the magnetisation value reported by Granado et al. \cite{Granado2001} can be misleading as there is no a priori way to determine the value of $A$ in Eq. (2). This leaves the exponential term factorisable with two free prefactors as, 
\begin{equation}
A \exp\bigg[-\frac{T-T_N}{T_0}\bigg] = A \exp\bigg[\frac{T_N}{T_0}\bigg] \exp\bigg[-\frac{T}{T_0}\bigg].
\end{equation}
As there is no way to fix the value of $A$, an empirical parameter, the term $\exp(T_N/T_0)$ serves merely as a scaling factor and does not influence the behaviour and form of the function with respect to $T$. Only the product $A\exp(T_N/T_0)$ is optimised in any fit, leaving the value of $T_N$ free as long as $A$ is scaled appropriately. Further, this form of a linewidth function does not diverge at $T=T_N$, which is contrary to an EPR description of the linewidth near an AFM transition, where the signal is expected to disappear \cite{Martinho2001,Hemmida2017} and hence $\Delta H$ $\rightarrow$ $\infty$. This suggests that the success of this description and the fit of Eq. (2) for the CaMnO\textsubscript{3} data yeilding $T_N$ equal to the magnetisation value as reported by Granado et al. in \cite{Granado2001} is perhaps misleading. We therefore dismiss this form of the linewidth as a useful reflection of the physics of this system which is known to have a clear AFM transition probed by both heat capacity and magnetisation measurements \cite{Neumeier2001,Cornelius2003}. Using the values obtained from the BKT fit and $T_N$ as 130 K as reported in \cite{Cornelius2003} in Eq. (5), we get a $J/J'$ value of the order of 1, which is consistent with the notion of CaMnO\textsubscript{3} as a 3D manganite. 

From Figure 1 and the $R^2$ values in Table 1 for LCMO, it is clear that the critical model and BKT model are both equally satisfactory in describing the divergence of the $\Delta H$ on approaching the transition to an AFM phase. This is understood by noting that an approximation of the theoretical expression for the BKT transition, i.e, Eq. (4) by critical behaviour yields $p \leq 3b/2 \simeq 2.4$ \cite{Heinrich2003}. Furthermore, the critical exponent is indicative of the dimensionality of the fluctuations and theoretical calculations have predicted that for the 3D Heisenberg AFM p = 1.7, for the 3D Ising AFM p = 1.8, and for the 2D Ising AFM p =3.3 \cite{Benner} - all of which differ from our experimental value of 2.8. Our value, $p=2.8$ is close to the experimentally reported critical exponent $p=2.6$ in several 2D magnets all of which are considered to be good realisations of weakly anisotropic layered 2D Heisenberg antiferromagnets \cite{Benner,Heinrich2003}. This is supported by the $J/J'$ value found from Eq. (5) using $T_N$ = 95K as reported in \cite{Granado2001}, which gives $J/J' \simeq 10^{3}$, which is of the order typically reported for weakly anisotropic layered 2D Heisenberg antiferromagnets \cite{Heinrich2003}. We conclude that this is strong evidence for the onset of some kind of two dimensionality in LCMO which appears to behave as a weakly anisotropic layered 2D Heisenberg antiferromagnet.

In the case of BSMCO, Bhagyashree et al., \cite{Bhagyashree} provide evidence for the coexistence of AFM and FM phases. Even for the undoped parent compound BSMO, Hervieu et al. report the presence of charge ordered Mn\textsuperscript{3+} and Mn\textsuperscript{4+} stripes which on doping with Cr\textsuperscript{3+} could conceivably lead to spin clusters \cite{Hervieu2001}. Here, the critical and BKT model both appear to describe the transition equally well. However, the value of the critical exponent is significantly smaller than any theoretically predicted value\cite{Benner}. The computed $J/J'$ values for BSMCO are of the order of 1 suggesting a lack of any two dimensionality and the BKT fit yields a low transition temperature of 14.7 K.

To understand the surprisingly good fit of the BKT transition to the data in all three 3D manganites we suggest that in all samples, especially BSMCO and CaMnO\textsubscript{3} where $J/J' \simeq 1$, it is the externally applied magnetic field as part of the EPR experiment that contributes to the onset of planar anisotropy. It does this by disallowing out of plane deformations along the applied field axis, which leads to XY behaviour and a BKT-like transition at a characteristic temperature, $T_{BKT}$. Vortex-like topological defects required for a BKT transition cannot be removed by moving spins out of the plane without a considerable free energy cost from the Zeeman energy. Experimental and quantum monte carlo simulation based reports of field induced BKT transitions in layered 2D Heisenberg AFMs argue that the applied field breaks the O(3) symmetry in the Heisenberg model reducing it to an O(2) symmetry which gives rise to genuinely XY behaviour of the spins over an extended temperature region \cite{Cuccoli2003,Baranov2016}. To quantify the competition between temperature related fluctuations and the applied magnetic field induced anisotropy energy, we introduce a dimensionless constant $\eta$,
\begin{equation}
    \eta = \frac{[\mu \cdot H] + i} {k_B T} 
\end{equation}
where $i$ is some measure of an intrinsic anisotropy energy which is linked to the sample's magnetic, structural and doping anisotropies, $H$ is the applied magnetic field of the EPR experiment and $k_BT$ represents the thermal fluctuation energy. When $\eta \gg 1$, then two dimensionality of the sample is induced as the thermal fluctuations along the applied magnetic field axis are much smaller than the magnetic interaction and intrinsic anisotropy energy and when $\eta \ll 1$, the sample is isotropic as thermal fluctuations dominate. 
\begin{figure}
\centering
{%
  \includegraphics[width=0.45\textwidth]{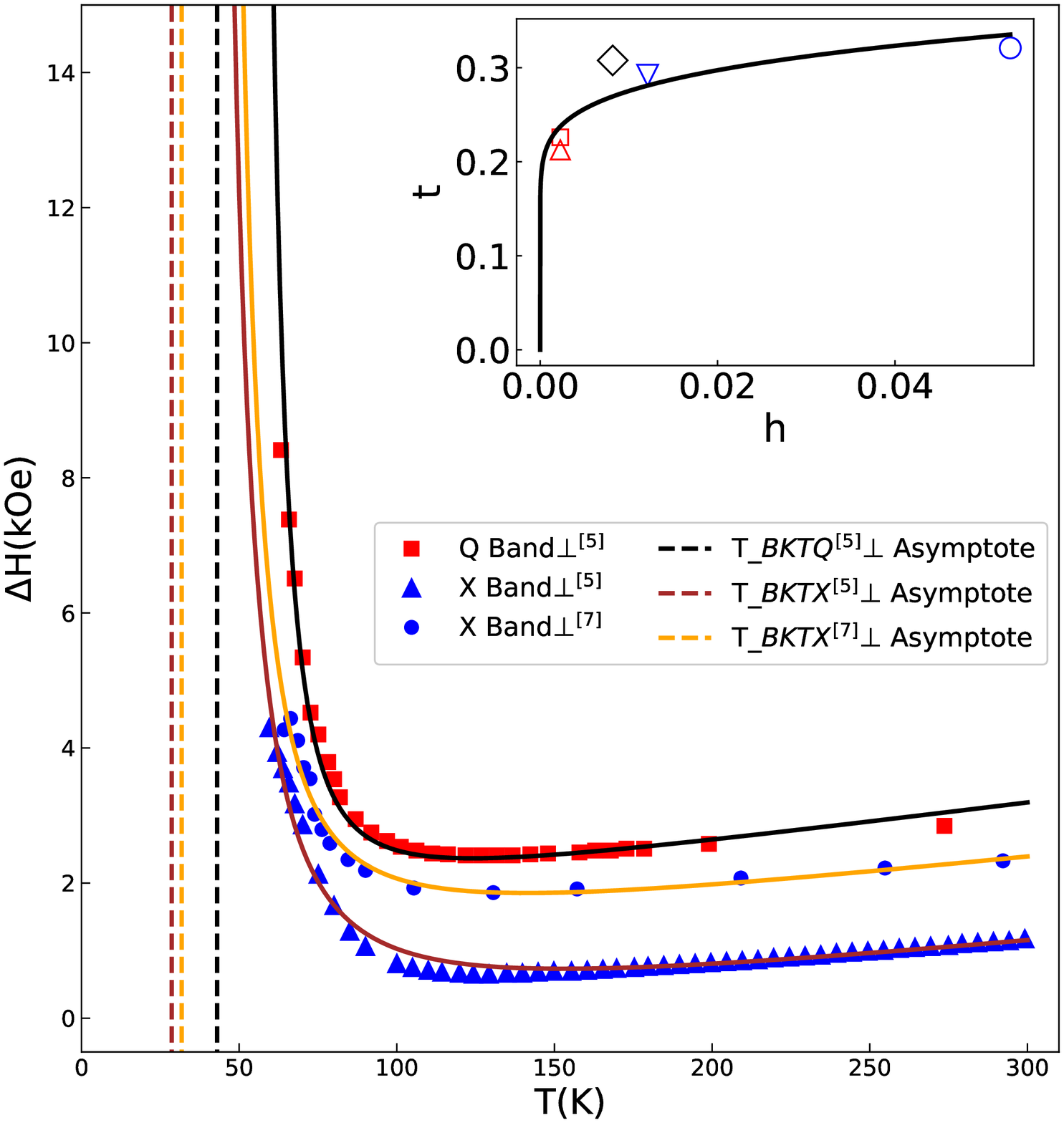}%
  }
\caption{X-Band($\bot c$) $\Delta H$ Fits for BaNi\textsubscript{2}V\textsubscript{2}O\textsubscript{8} reported in \cite{Heinrich2003} and \cite{Waibel} and Q-Band($\bot c$) $\Delta H$ Fits for BaNi\textsubscript{2}V\textsubscript{2}O\textsubscript{8} reported in \cite{Heinrich2003}. Inset: Eqn (6) predicted field dependence fit of Q band (1.08 T)($\Diamond$), X band(.30 T)($\triangle$) EPR $t_{BKT}$ from \cite{Heinrich2003}, X band(.30 T $\square$) EPR $t_{BKT}$ from \cite{Waibel} and NMR extracted $t_{BKT}$ at 1.6 T ($\nabla$) and 7.0 T ($\bigcirc$) from \cite{Waibel}.}
\end{figure}

At $T=T_{BKT}$ from the fits, H = 0.335 T (value of resonance field in X band for LCMO, CaMnO\textsubscript{3} and BSMCO) and with $i=0$ we find that $\eta \simeq 0.1$ for both CaMnO\textsubscript{3} and BSMCO \cite{Granado2001,Bhagyashree}. We interpret this as an indication that the onset of two dimensional and consequent BKT-like behaviour in these samples occurs when the field interaction energy is close to the order of the thermal energy. For the same applied field in LCMO \cite{Granado2001}, we find that $T_{BKT}$ is much larger, implying a large, non zero value for $i$ in LCMO assuming $\eta$ to be universally indicative of a BKT onset threshold. We attribute this large intrinsic anisotropy in LCMO to the presence of 5\% La doping. Doping is known to result in inhomogeneous magnetism and phase separation consisting of G-AFM background and FM clusters \cite{Ling}. This work concludes that as a function of doping level the FM double exchange interaction changes dimensionality from 0D to 3D. Evidence is also available for the presence of nanometric scale spin clusters in LCMO \cite{Granado2003} which are a manifestation of this intrinsic anisotropy and could serve as vortices during the BKT transition \cite{Granado2003}.

To further test our hypothesis that the applied EPR field is responsible for the onset of BKT-like behaviour in manganites, we predict that a higher EPR field allows for a higher temperature onset of the threshold anisotropy required for BKT behaviour which will manifest as a higher $T_{BKT}$ as predicted by both Eq. (6) and Eq. (8). We consider for an example, results of Heinrich et al. of $\Delta H(T)$ \cite{Heinrich2003} probed in X (9.4GHz) and Q (34GHz) bands for BaNi\textsubscript{2}V\textsubscript{2}O\textsubscript{8} (BNVO) as the data is available for significantly different applied fields. BNVO is not a manganite but can be used to test the hypothesis which does not implicitly require the sample to be a manganite.

For both sample orientations ($H\bot c$ and $H\parallel c$), we have fit the Q and X band probed $\Delta H(T)$ to Eq. (4) with $b=\pi/2$. From Figure 3 it is immediately apparent that the X band $\Delta H(T)$ diverges at a lower temperature for the $\bot c$ data and we have found consistent results for the $\parallel c$ $\Delta H(T)$ data. These fits yield $T_{BKT_{Q\bot}}$ = 43.3 K and $T_{BKT_{Q\parallel}}$ = 40.3 K for the Q-band (the same as reported by Heinrich et al.) and $T_{BKT_{X\bot}}$ = 29.9 K and $T_{BKT_{X\parallel}}$ = 34.8 K for the X-Band. Heinrich et al. suggest that their X-Band data may not be reliable as the resonance field value exceeds the linewidth at low temperatures. In view of this we fit X-band EPR $\Delta H(T)$ data from a more recent study by Waibel et al. on the same compound (also plotted in Figure 3). From their X-Band $\bot c$ data we find $T_{BKT}$ = 31.8 K which is consistent with our analysis of the X band data from Heinrich et al. and again significantly lower than the value for Q-band in \cite{Heinrich2003}. 

Using the reported value of $g_\bot = 2.243$ for Ni\textsuperscript{2+} system \cite{Heinrich2003}, the magnetic field at the Q and X band were calculated to be $H_Q = 1.08 T$ and $H_X = 0.30 T$ respectively, confirming our hypothesis that an increase in the applied field results in an increase in the threshold anisotropy onset temperature and consequently $T_{BKT}$. The $J/J'$ values calculated (using $T_N$ = 50 K reported in Inelastic Neutron Scattering measurements) \cite{Rogado2002}, for this sample are significantly field dependant, with the Q-Band($\bot c$) $J/J' \simeq$ 3000 and the X-Band($\bot$ c) $J/J' \simeq 40$, a large decrease. This further confirms the dependence of the planar magnetic anisotropy on the value of the applied magnetic field. 

We calculate the $i$ from Eq. (8) for both $\bot c$ and $\parallel c$ for BaNi\textsubscript{2}V\textsubscript{2}O\textsubscript{8} to be of the order of an electron's magnetic interaction energy. We also note that $i_{\parallel c} \simeq 3i_{\bot c}$. We fit Eq. (6) to calculated $t_{BKTX\bot}$ $t_{BKTQ\bot}$ and their respective $h_X$ and $h_Q$ as well as NMR longitudinal relaxation time signalled BKT transition temperatures at 1.6 and 7 T extracted from data reported in \cite{Waibel} using the nearest neighbour magnetic exchange interaction energy for BaNi\textsubscript{2}V\textsubscript{2}O\textsubscript{8} reported by Klyushina et al.($J = 12.125 meV$) \cite{Klyushina2017}. We find the spin stiffness $\rho_S = 0.408 J$ from our fit in Figure 3(inset), which is the same order of magnitude as the $S=1/2$ isotropic Heisenberg AFM value of $0.18 J$ reported in \cite{RRPSingh}. We suspect that more $(h,t)$ points and including $J_{nn}$ and $J_{nnn}$ interactions will lead to a more precise $\rho_S$ value for this $S=1$ lattice, as the spin stiffness value extracted is very sensitive to the number of data points. 

In summary, we have attempted to explain the fit of the BKT model in these 3D samples by showing that the energy scales of the EPR experiment's own applied magnetic field combined with the sample's intrinsic anisotropy energy $i$ is comparable to the thermal energy at $T_{BKT}$. This suggests that the EPR experiment's applied magnetic field can lead to planar anisotropy and genuinely XY behaviour by energetically disallowing the removal of vortex-like topological defects at $T\leq T_{BKT}$. This relationship between $T_{BKT}$ and the applied field is even seen in BaNi\textsubscript{2}V\textsubscript{2}O\textsubscript{8}, a quasi-2D AFM with a calculated high intrinsic planar anisotropy and a consequently low anisotropic contribution from the applied field. We conclude that H has significant influence even when intrinsic 2-dimensionality is present in the system and thus provides tunable XY behaviour in magnetic insulators \cite{Baranov2016,Hemmida2017}.We suggest that similarly high intrinsic anisotropy, due to 5\% La doping in CaMnO\textsubscript{3}, causes the high temperature onset of quasi two dimensionality in an otherwise latent (low-temperature) XY sample: undoped CaMnO\textsubscript{3}. 

\begin{acknowledgments}
We gratefully acknowledge useful discussions with H.R. Krishnamurthy (Indian Institute of Science, Bangalore). SVB and AA thank the Indian Academy of Sciences (IAS), National Academy of Sciences, India (NASI) and Indian National Science Academy (INSA). 
\end{acknowledgments}

%\bibliographystyle{apsrev4-1} 
%\bibliography{bibl}

%\input{Bib_PRL.bbl}


%

\end{document}